# Online Delay Estimation and Adaptive Compensation in Wireless Networked System: An Embedded Control Design

Santosh Mohan Rajkumar, Sayan Chakraborty, Rajeeb Dey, and Dipankar Deb

**Abstract:** In the present work, an embedded PI controller is designed for speed regulation of DC servomotor over a wireless network. The embedded controller integrates PI controller with a proposed time-delay estimator and an adaptive digital Smith predictor for its real time operation, which is the novelty of the work. The real time or rather online operation of the developed embedded controller over wireless network is possible mainly due to the contributions in terms of proposing a new empirical formula for time-delay estimation, viable approximation of time-delay, discretization scheme in developing the proposed adaptive Smith predictor. The proposed embedded design validates that the deterioration of control performance due to random network-induced delay are mitigated in real time. The embedded controller is designed and tested for a short-range of 100 meters, and wireless technology chosen is suitable for biomedical applications.

## 1. INTRODUCTION

In medical and industrial applications, there is an onset of real-time control over the wireless network in recent times. Networked control system (NCS) interconnects the sensors, actuators, and controllers through wired or wireless networks [1] with potential industrial applications with rapid development of data communication network technologies. Merits of network technologies are easy maintenance and expandability for the control system design, with certain drawbacks like time-delay, packet dropout, and message collision. Focus areas of research in NCS includes improved packet dropout, time-delay, jitter, channel capacity (or) throughput, network scheduling [2–4], communication protocols [5, 6] and design of control algorithms [1, 7–10].

A major concern of NCS is the stability of the physical system and performance degradation due to uncertain time-delay from the network serving the control application. Also, in the presence of time-delay, the system exhibits transcendental characteristics [11] wherein system analysis and controller design is challenging. The design of control architecture for NCS in the presence of uncertain time-delays is a popular area of research. The main focus of the present work is to develop an embedded control architecture consisting of a time-delay estimator and a time-delay compensator. Over the past years, various methodologies have been proposed by researchers in this area. Some of the most relevant ones are addressed below. An adaptive gain output feedback $H_\infty$ controller is designed in [10] to control the effects of time-delays and packet losses arising in NCS. The performance of the proposed methodology was evaluated on a temperature control plant with network-feedback loop. A robust Smith predictor to compensate round trip time-delay in NCS setting [4] shows that stability is guaranteed up to an upper bound of round trip delay. Online time-delay estimation using round trip time measurement (RTT) followed by an adaptive Smith predictor scheme in NCS setting is studied in [1]. To test feasibility, an AC 400W motor was controlled from a distance of 15km. An improved predictor control strategy for data dropout and time-delay in NCS is proposed in [12], and a model-based prediction control methodology is proposed in [13] where the current and predicted control output is calculated for several time steps and sent to the actuator node at once. A predictive NCS [14] is particularly active to compensate network delays and data dropouts. Other model predictive control methodologies [15, 16] are proposed for similar scenarios are given in references therein. Studies on convergence analysis of NCS are found in [17,18]. In the present work, a linear plant model is considered. Non-linear systems in presence of time-delays can be found in [19, 20].

NCS exhibits stochastic behavior due to packet dropout, time-delay, etc.. Recent developments in the stochastic networked-based framework are found in [21,22], and references therein. Significant contributions have been made to overcome issues like uncertain delays, packet drop-outs, etc.. Robust stability analysis of NCS with time-varying delays in LMI framework is reported in [23] in a polytopic framework which also discusses the packet drop-out cases. In [10] an experimental result for temperature control over the wireless network is reported where the control gain and the stability margins are computed in an LMI framework. If the system model is expressed in a nonlinear state-space form then, one can use the fuzzy T-S modeling approach as in [24] to obtain a polytopic system representation. The polytopic linear models so obtained allow designing NCS with state as well as output-feedback control law as in [10, 24] in an LMI framework. This control design requires a state space system model which is not the case in the present work wherein our prime focus is on uncertain time-delay arising in NCS. While several methodologies to overcome uncertain delay has been studied in simulation, the issue of practical implementation has not been explored much. [1, 10] provide experimental validation with the controller implemented on a PC which is costly, and the architecture is not portable. The present work is developed for easy and cost-effective embedded implementation, and overall portable architecture uses wireless communication that provides flexibility. The development of NCS in the present work is demonstrated for a medical application (artificial pancreas system (APS) [25]) where connectivity between devices is restricted to a limited region. APS is crucial for Type 1 Diabetic patients whose pancreas does not secrete enough insulin thereby leading to the risk of hyperglycemia (rise in glucose level) and hypoglycemia (sudden fall in glucose level) due to an improper insulin dosage [26]. There is a need for tight glycemic control, monitoring, and management of glucose level in real-time for these patients. In APS, there exists sufficient short-range wireless communication between glucose sensor, insulin infusion pump, control algorithm on embedded computers [27]. The reliability of the APS system can be enhanced by incorporating continuous monitoring facilities by doctors/nurses located at large/medium distances from the patients. Bluetooth piconet or scatternet technology is used in this work for the connectivity of devices over short or medium range.

The main aim of the paper is not to develop an APS but to develop a Wireless Network Control System (WNCS) based actuation control system in the face of network induced uncertain time-delay. However, the developed experimental environment for WNCS based DC servo motor speed control using Bluetooth technology in an embedded platform (Arduino) can be easily adapted for the APS as the arrangement of the devices are similar. The main contributions of this work are (i) validation through experimental results on how time-varying or random network delay affects actuation system stability and performance, (ii) online delay estimator design with an adaptive Smith predictor for mitigating uncertain random time-delay, (iii) deploying the entire computation in an embedded platform unlike in [10] and references therein. For the implementation of delay estimation, adaptive Smith predictor and digital controller algorithms, certain computational modifications are proposed.

The paper is organized as follows. Section 2 presents the experimental setup and the control architecture of WNCS. Section 3 discusses measurement of round trip time (RTT) and time-delay estimation. Section 4 discusses the proposed adaptive digital Smith predictor scheme. Section 5 discusses a posteriori stability analysis of the time-delayed DC motor system sans Smith predictor. Section 6 discusses the experimental results, Section 7 provides algorithmic steps of the developed time-delay compensator, and Section 8 provides concluding remarks.

## 2. EXPERIMENTAL SETUP

This section discusses the experimental setup of WNCS using Bluetooth technology for connecting a limited number of devices for voice and data messages over a short range but has an obvious advantage in medical applications as it operates in ISM band of 2.4 GHz. The basic unit of a Bluetooth network is a 'piconet' as shown in Fig. 1. Piconets are expandable to a bigger network called a 'scatternet' with many master bridges as shown in Fig. 2. Sufficient uncertain transmission delay may occur between different piconets when a bridge is serving a particular piconet and master in another piconet is waiting for the current service [28]. Uncertain delays may affect system stability and performance and are the motivation of the present work wherein a conceptual arrangement of devices in a reliable artificial pancreas system (APS) operating with Bluetooth technology is shown in Fig. 3. APS is an automated system that mimics the homeostasis process to regulate plasma glucose in the human body.

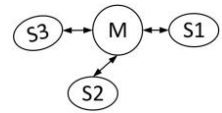

Fig. 1. Bluetooth piconet. (M=Master, S=Slave).

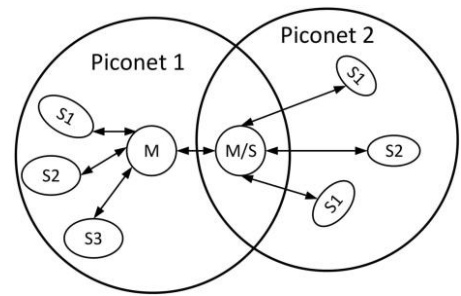

Fig. 2. Bluetooth scatternet.

This automation is depicted in a closed-loop architecture with wireless network technology in Fig. 4.

To realize an automation system similar to Fig. 4, a bluetooth enabled DC servo motor speed control is developed, and the physical set up is shown in Fig. 5.

The developed prototype operates with a single piconet. However, to realize a massive transmission delay due to the scatternet formation, the master node of the piconet (intermediate node) runs an algorithm to conceptually emulate uncertain random delay in the network [28]. The various hardware components used for prototyping the proposed experimental setup are described below:

A permanent magnet type DC motor is used with 200 rps rated speed and 3V-5V rated voltage. For controller design, the transfer function of the DC motor is identified using the system identification technique and is given as

$$G(s) = \frac{4.159}{s + 3.888}. \quad (1)$$

A discrete time equivalent of this plant with a sampling time of 0.02s can be obtained as

$$G(z) = \frac{0.0831}{z - 0.92}. \quad (2)$$

The speed sensor module is an HC-020k optical speed encoder with a measurement frequency of 100 KHz, encoder resolution of 20 lines, and working voltage range of 4V-5V. As the motor shaft turns, the slots pass the LED - photo-transistor pairs creating a series of electrical ON/OFF states. For every electrical ON and OFF state, a binary data 1 and 0 is fed to the embedded microcontroller to determine the motor speed.

The motor driver used is a 16 pin Texas Instrument L293D driver IC with a supply voltage range of 4.5 V-36 V, output current 600 mA per channel, peak output current 1.2 A per channel and speed controller input PWM pins. The Bluetooth device is a V2.0+EDR (Enhanced Data Rate) HC05 Bluetooth Serial Port Protocol (SPP) module with 3 Mbps data rate, 2.4 GHz radio transceiver.

The implementation of the digital controller, delay estimator, and adaptive Smith predictor are done on Arduino UNO and Arduino DUE. For digital signal processing in the control block, inbuilt pulse width modulator (PWM) of Arduino is used. A discrete PI controller is implemented in embedded hardware, as

$$u(n) = K_P\, e(n) + K_I \sum_{i=0}^{n-1} e(i). \quad (3)$$

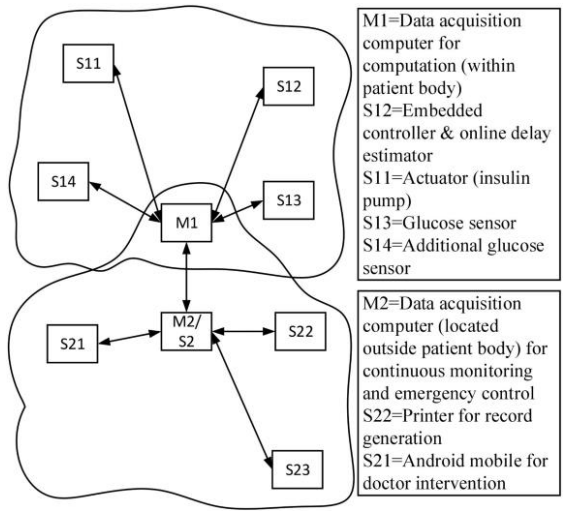

Fig. 3. Proposed scatternet based reliable APS (M=Master, S=Slave).

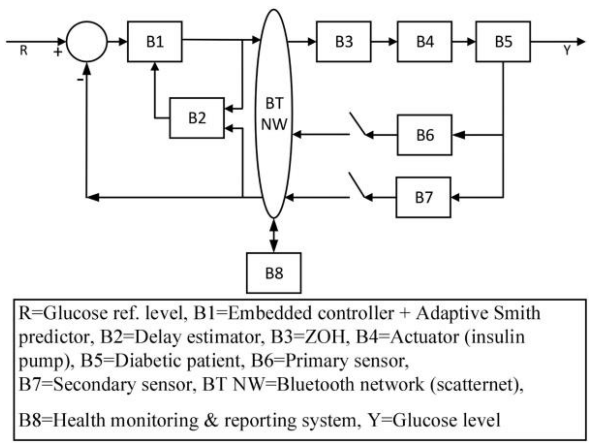

Fig. 4. Proposed wireless network control for the APS.

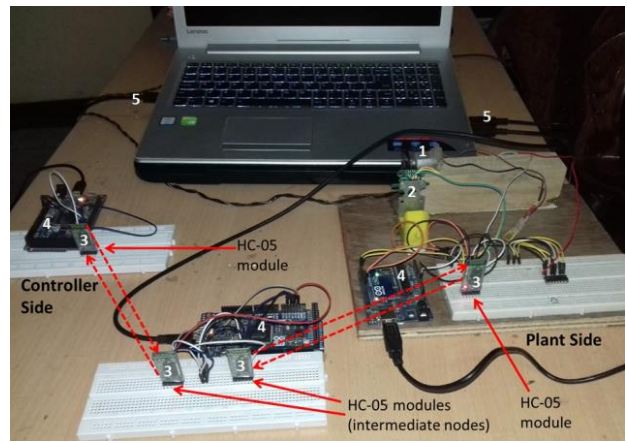

Fig. 5. Physical setup of the experiment.

This form of PI controller is known as the position algorithm. The main concern with the position algorithm is that all the past error values need to be stored. While working with an Arduino microcontroller, this problem is addressed by assigning an integer variable for the summation of the error signal. An integer in the microcontroller stores a 32-bit value that yields a range of $-2,147,483,648$ to $2,147,483,647$ which is sufficiently large for the present application. The tuning of the PI controller parameters is done using the SISO toolbox of MATLAB. The obtained controller gains are $K_P = 1.69$ and $K_I = 0.1488$.

Fig. 6 shows the block diagram of the proposed control system. The two embedded microcontrollers, Arduino board 1 (AB1) and Arduino board 2 (AB2), communicate through a wireless network viz. Bluetooth. AB1 and AB2 are individual 'piconets', both capable of connecting to 7 different slaves each. A third microcontroller, Arduino board 3 (AB3) is an intermediate node between AB1 and AB2. The only communication channel between AB1 and AB2 is the intermediate node AB3, thereby forming a 'scatternet' like structure. The output signal from AB3 is the randomly delayed signal received either from AB1 or AB2. Such a configuration emulates a similar situation of uncertain delay in a scatternet Bluetooth network.

AB1 consists of a time-delay estimator, a time-delay compensator and a PI controller. The inbuilt PWM in AB1 converts the controller output to an 8-bit PWM signal. This signal is then transmitted to AB2 via the Bluetooth network of AB3. The motor driver IC, DC motor and the speed sensor are interfaced with AB2. The motor driver IC generates a voltage based on the duty ratio of the PWM signal received from AB1. The generated voltage is then given as an input to the DC motor. The speed sensor block in Fig. 6 measures the speed of the DC motor. The measured speed is then transmitted to AB1 via the Bluetooth network of AB3. AB3 introduces random time-delays to the PWM signal and the measured speed signal before transmitting them to AB2 and AB1 respectively. Hereafter, the delayed speed signal is referred to as the delayed feedback signal. A send time stamp is recorded in AB1 for every PWM signal sent to AB2 from AB1 and a receive time stamp is recorded in AB1 for every delayed feedback signal received in AB1 from AB2. The output of the comparator 3 in Fig. 6 is the difference between the time stamps and is referred to as the round trip time (RTT). Based on the RTT measured and the delayed feedback signal received, the online time-delay estimator estimates the delay which is then fed to the online time-delay compensator. The output of the time-delay compensator is compared by comparator 2 with the error signal generated from comparator 1. The output of comparator 2 is the non-delayed nominal output (see (13)) in Section 4.1.4) which is then fed to the discrete controller for the desired action for the DC motor.

## 3. RTT MEASUREMENT & DELAY ESTIMATION

Communication delay is introduced by the Bluetooth network while the controller and plant node communicate via an intermediate Bluetooth network as depicted in Fig. 6. Further, to emulate a random delay, the messages received to and from either controller or plant are randomly delayed by the intermediate node. Uncertain delay in WNCS degrades performance and system stability under control [10]. In the developed prototype, the time-delay induced from the controller-to-plant direction is taken as $t_1$ (denoted as forward delay) and the time-delay from the sensor-to-controller direction is $t_2$ (denoted as backward delay). Total time-delay $t_d = t_1 + t_2$ is the round trip time (RTT).

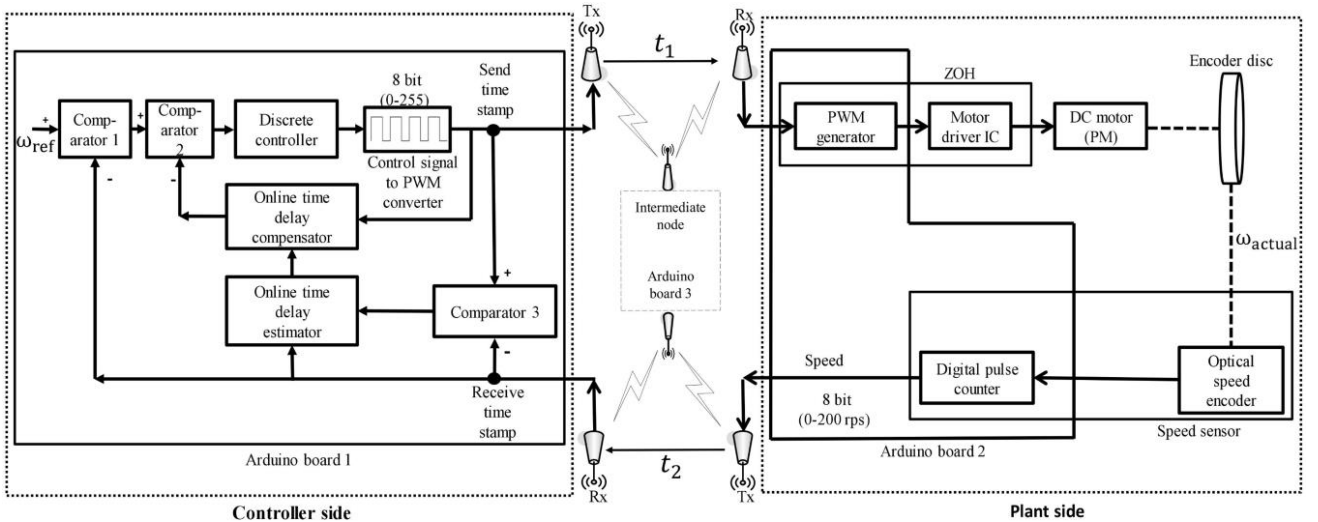

Fig. 6. Practical wireless block diagram for speed control of DC motor.

### 3.1. RTT measurement

For RTT measurement and estimation of time-delay, the following rules are adopted from [1].

- **Normal transmission:** If $t_d$ is less than one sampling period, the effect on the system performance is negligible, and the measured $t_d$ is the estimated delay ($t_m$).
- **Vacant sampling:** In the absence of occurrence of a data sample before the advent of the next sampling instant, the $t_d$ measured before is added to one sampling period and is considered as $t_m$.
- **Message Rejection:** If there is occurrence of more than one data at a sampling instant, the most recently arrived data is accepted and all the previous data are discarded. Accordingly, the recently measured $t_d$ is considered as $t_m$.
- **Delayed transmission:** The $t_d$ measured continuously is considered as $t_m$.

In the developed prototype, plant side is considered to be time-driven (or clock-driven) while the controller side is considered to be event-driven. A graphical illustration of RTT measurement and time-delay estimation on the proposed setup is presented in Fig. 7. The time-delay estimation data is tabulated in Table 1. The occurrence of zero is considered equivalent to the vacant sampling situation.

### 3.2. On-line delay estimation

The proposed empirical formula for time-delay estimation based on the RTT measurement is expressed as

$$t_m = t_{2pr} - t_{1pr} + \sum_{i=2}^{v_s} t_{pai}, \quad (4)$$

Table 1. Illustration of delay estimation.

| Sent data | $t_{1pr}$ | $t_{2pr}$ | $t_{pa}$ | Received data | $t_m$ | Sample time |
|---|---|---|---|---|---|---|
| 50 | 0 | 23 | 23 | 0 | 0 | 0 |
| 70 | 23 | 45 | 22 | 0 | 20 | 20 |
| 80 | 45 | 74 | 29 | 50 | 40 | 40 |
| 90 | 74 | 93 | 19 | 70 | 60 | 60 |
| 100 | 93 | 108 | 15 | 80 | 74 | 80 |
| 110 | 108 | 124 | 16 | 90 | 70 | 100 |
| 120 | 124 | 143 | 19 | 100 | 63 | 120 |
| 130 | 143 | 167 | 24 | 110 | 50 | 140 |
| 140 | 167 | 184 | 17 | 120 | 50 | 160 |

where $t_m$ is the estimated delay, $t_{1pr}$ is the present sent time-stamp, $t_{2pr}$ is the present receive time-stamp, $t_{pa}$ is the difference of past time-stamps, $v_s$ is the number of vacant samples. The computation of estimated delay using (4) is explained with an example. It is evident from Table 1 that $v_s = 3$. Considering the estimated delay ($t_m$) at sample time 100 ms, the present time stamps are $t_{2pr} = 93$ ms, $t_{1pr} = 74$ ms, and the difference of the past time stamps are $t_{pa2} = 29$ ms, $t_{pa3} = 22$ ms. Using (4), we get $t_m = (93 - 74) + 29 + 22 = 70$ ms. Similarly, at sample time 140 ms, $t_m = (124 - 108) + 15 + 19 = 50$ ms. Note that, here we consider $t_{2pr} - t_{1pr}$ as the RTT which is same as $t_1 + t_2$.

In the proposed prototype, the measurement and estimation of delay are done at the controller side (see Fig. 6) thus making the implementation easier and faster. Further, unlike in [1], the measurement and estimation is implemented in a dedicated embedded platform and not in a high-performance computer, and so a simple yet accurate algorithm needed to be developed.

### 4. TIME-DELAY COMPENSATION

The closed-loop transfer function of a delayed system along with the classical Smith predictor as seen from Fig. 8(a) is found to be

$$\frac{Y(s)}{R(s)} = \frac{G_s(s)G(s)e^{-t_1 s}}{1 + G_s(s)G(s)e^{-t_d s}}, \quad (5)$$

where, $Y(s)$ is the output, $R(s)$ is the input/reference signal, and $G_s$ is the inner loop transfer function given as

$$G_s(s) = \frac{G_c(s)}{1 + (1 - e^{-t_m s})\hat{G}(s)G_c(s)} \quad (6)$$

Substituting (6) in (5), we get

$$\frac{Y(s)}{R(s)} = \frac{G_c(s)G(s)e^{-t_1 s}}{1 + \hat{G}(s)G_c(s)(1 - e^{-t_m s}) + G(s)G_c(s)e^{-t_d s}}, \quad (7)$$

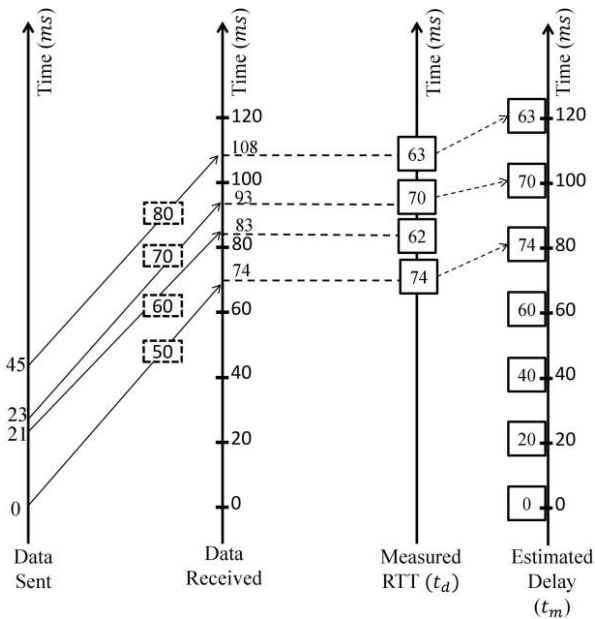

Fig. 7. Measured RTT ($t_d$) and estimated delay ($t_m$).

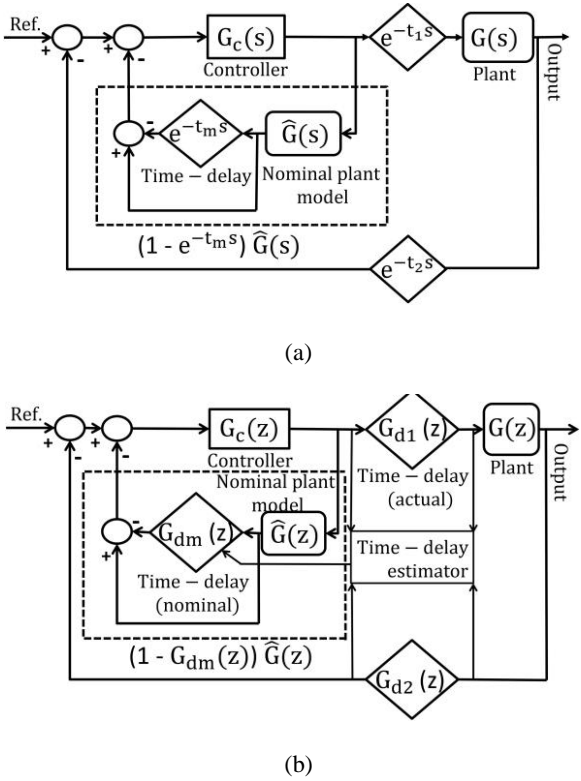

Fig. 8. Block diagrams for (a) Classical Smith predictor; (b) Digital Smith predictor.

If $t_m = t_d$ and $\hat{G}(s) = G(s)$ in (7), then it reduces to

$$\frac{Y(s)}{R(s)} = \frac{G_c(s)G(s)}{1 + G_c(s)G(s)} e^{-t_1 s}. \tag{8}$$

The effect of time-delay is compensated from feedback signal and the controller takes a decision sans any delayed information. Classical Smith predictor is useful when the system time-delay is fixed. But in practice, the network-induced delay may vary due to which actual delay differs from the estimated value, and so an adaptive version of the Smith predictor using delay estimator is needed.

### 4.1. Synthesis of adaptive digital Smith predictor

The focus of present work is to implement an adap- tive Smith predictor in dedicated embedded hardware with limited memory and computing power. An adaptive Smith predictor scheme adopted from [1], is shown in Fig. 8(b).
The transfer function of the adaptive Smith predictor is

$$G_{sp}(s) = (1 - e^{-t_m s})\hat{G}(s). \tag{9}$$

Due to the $e^{-t_m s}$ term, the transfer function has infinite roots. In the present work, for the first time, digital implementation of the adaptive Smith predictor scheme in an embedded platform is proposed, and the procedure adopted is outlined below.

Table 2. Different delay approximation series

| Series | Formulation |
|---|---|
| Pade | $\dfrac{1 - 0.5s\tau + 0.0833s^2\tau^2}{1 + 0.5s\tau + 0.0833s^2\tau^2}$ |
| Marshall | $\dfrac{1 - 0.0625s^2\tau^2}{1 + 0.0625s^2\tau^2}$ |
| Product | $\dfrac{1 - 0.5s\tau + 0.125s^2\tau^2}{1 + 0.5s\tau + 0.125s^2\tau^2}$ |
| Laguerre | $\dfrac{1 - 0.5s\tau + 0.0625s^2\tau^2}{1 + 0.5s\tau + 0.0625s^2\tau^2}$ |
| Paynter | $\dfrac{1}{1 + s\tau + 0.405s^2\tau^2}$ |
| Direct frequency response (DFR) | $\dfrac{1 - 0.49s\tau + 0.0954s^2\tau^2}{1 + 0.49s\tau + 0.0954s^2\tau^2}$ |

Table 3. Time-delay approximation errors.

| Series | $\tau = 0.04$ | $\tau = 0.24$ | $\tau = 1$ | Average ISE |
|---|---|---|---|---|
| Pade | 0.0057 | 0.0345 | 0.1437 | 0.0592 |
| Marshal | 43.989 | 41.139 | 41.880 | 41.321 |
| Product | 0.0054 | 0.0321 | 0.1339 | 0.0552 |
| Laguare | 0.0068 | 0.0406 | 0.169 | 0.0696 |
| Paynter | 0.0058 | 0.0347 | 0.1447 | 0.0596 |
| DFR | 0.0053 | 0.0319 | 0.1331 | 0.0548 |

#### 4.1.1 Time-delay approximation

Approximate $e^{-t_m s}$ using a suitable series polynomial as given in Table 2 such that $e^{-t_m s} \approx G_{dm}(s)$. The occurrence of an error in the approximation is inevitable. The error also depends on the type and order of the series being considered for approximation.

Approximation errors are tabulated in Table 3 for different types of series polynomials. The DFR and Pade series gives the minimum average error. In this work, we use DFR and Pade series for time-delay approximation. However, the experimental results are shown for the DFR series approximation only.

#### 4.1.2 Discretisation

Discretise the transfer function $G_{sp}(s) = (1 - e^{-t_m s})\hat{G}(s) = (1 - G_{dm}(s))G(s)$ using a suitable transform and obtain $G_{sp}(z) = (1 - G_{dm}(z))G(z)$. The nominal plant is selected to be a discrete version of the plant given by $\hat{G} \; z\frac{0.0831}{z-0.92}$. Approximation of $e^{-t_m s}$ using DFR series approximation as indicated above gives $G_{dm}(s)$. Further, using the bilinear transform with sampling time $T = 20$ms, the discrete transfer function of $G_{dm}(s)$ is obtained as

$$G_{dm}(z) = \frac{c(t_m)z^2 + d(t_m)z + e(t_m)}{e(t_m)z^2 + d(t_m)z + c(t_m)}, \tag{10}$$

where $c(t_m) = 1 + 100t_m(9.54t_m - 0.49)$, $e(t_m) = 1 + 100t_m(9.54t_m + 0.49)$, $d(t_m) = 2 - 100t_m(9.54t_m + 0.49) - $

$100t_m (9.54t_m + 0.49)$. The coefficients of (10) are functions of the estimated delay. With algebraic manipulations in negative powers of $z$, one can get

$$G_{sp}(z) = \frac{Y(z)}{X(z)} = \frac{f(t_m)z^{-1} + g(t_m)z^{-3}}{1 + h(t_m)z^{-1} + i(t_m)z^{-2} + j(t_m)z^{-3}}, \quad (11)$$

where

$$f(t_m) = \frac{8.1438t_m}{954.0t_m^2 + 49.0t_m + 1.0} = g(t_m),$$
$$h(t_m) = \frac{-2785.68t_m^2 - 45.080t_m + 1.08}{954.0t_m^2 + 49.0t_m + 1.0},$$
$$i(t_m) = \frac{2709.36t_m^2 - 49.0t_m - 0.84}{954.0t_m^2 + 49.0t_m + 1.0},$$
$$j(t_m) = \frac{-877.68t_m^2 + 45.080t_m - 0.92000}{954.0t_m^2 + 49.0t_m + 1.0}.$$

Using Pade series, we get the following coefficients,

$$f(t_m) = \frac{24.93t_m}{2500.0t_m^2 + 150.0t_m + 3.0}, \quad g(t_m) = f(t_m),$$
$$h(t_m) = \frac{-7300t_m^2 - 138t_m + 3.24}{2500t_m^2 + 150t_m + 3},$$
$$i(t_m) = \frac{7100t_m^2 - 150t_m - 2.52}{2500t_m^2 + 150t_m + 3},$$
$$j(t_m) = \frac{-2300.0t_m^2 + 138.0t_m - 2.7600}{2500.0t_m^2 + 150.0t_m + 3.0}.$$

#### 4.1.3 Finding the difference equation

The difference equation from the discretised transfer function is obtained from (11) as

$$y[n] = f(t_m)x[n-1] + g(t_m)x[n-3] - h(t_m)y[n-1] - i(t_m)y[n-2] - j(t_m)y[n-3], \quad (12)$$

which is the output of the adaptive digital Smith predictor (time-delay compensator). This difference equation can be easily implemented on an embedded device for time-delay compensation.

#### 4.1.4 Effect of approximation error

The compensator output $y$ comprises of two parts: (i) the nominal feedback $\hat{y}$ by the nominal plant and (ii) delayed nominal feedback $\hat{y}_d$ by the nominal plant. From Fig. 6 one can see that the comparator 1 compares the reference signal $r$ with the delayed feedback $y_d$, following which the comparator 2 compares the output of comparator 1, $e_1$, with the compensator output $y$. Therefore, the output of the comparator 2 is obtained as

$$e_2 = e_1 - y = r - \hat{y} - (y_d - \hat{y}_d). \quad (13)$$

If the nominal plant model ($\hat{G}(z)$) and the time-delay approximation ($G_{dm}(z)$) are proper, the residue of $|y_d - \hat{y}_d|$ is negligible (converges to zero) and a delay free error signal is generated for proper controller action. In practice, the residue $y_d - \hat{y}_d$ can never be zero but there exists a small error $\varepsilon$. From Table 3 and [29], the approximation schemes like Pade and DFR provide reasonable approximation of the time-delay such that $|y_d - \hat{y}_d| \leq \varepsilon$. Hence, a satisfactory delay free error signal can be generated for proper controller action.

## 5. STABILITY ANALYSIS

In this section, the maximum analytical delay bound ($t_{crit}$) for the stability of the closed-loop time-delayed DC motor in the absence of Smith predictor is obtained.

The characteristic roots of the time-delay system are computed using the Galerkin approximation wherein the root convergence starts from the rightmost root within a specified tolerance $\xi$ [11, 30]. Stability is analyzed by studying the effect of delay on the movement of the rightmost characteristic roots. For brevity, the derivation of Galerkin approximation for computation of characteristic roots for time-delay systems (TDS) is not discussed here but can be found in [11].

The closed-loop transfer function of the DC motor system without the Smith predictor can be obtained as

$$G_{cl}(s) = \frac{G_c G e^{-t_1 s}}{1 + G_c G e^{-t_d s}}, \quad (14)$$

where $t_d = t_1 + t_2$, and the characteristic equation is

$$d(s) = s^2 + 3.888s + 7.0287 s e^{-t_d s} + 0.6188 e^{-t_d s} = 0. \quad (15)$$

The characteristic DDE of the system described in (14) can be easily obtained from (15) as

$$\ddot{x}(t) + 3.89\dot{x}(t) + 7.029\dot{x}(t - t_d) + 0.619x(t - t_d) = 0. \quad (16)$$

Applying Galerkin approximation theory, one can obtain the characteristic roots of (16) for different time-delays. The movement of the real part of the first four rightmost roots of (16) by varying $t_d$, is shown in Fig. 9. The dominant rightmost root (1st root) remains in the right half s-plane for delays greater than 0.369 s. Thus, the closed-loop system without Smith predictor loses its stability for delays greater than $t_{crit} = 0.369$ s (critical delay). Therefore, the main objective is to implement the proposed Smith predictor scheme such that the closed-loop system is stable for delays greater than $t_{crit}$.

## 6. EXPERIMENTAL RESULTS

The system responses of the following experimental setups are compared: 1) classical Smith predictor (CSP) with a PI controller and a fixed time-delay, and 2) adaptive

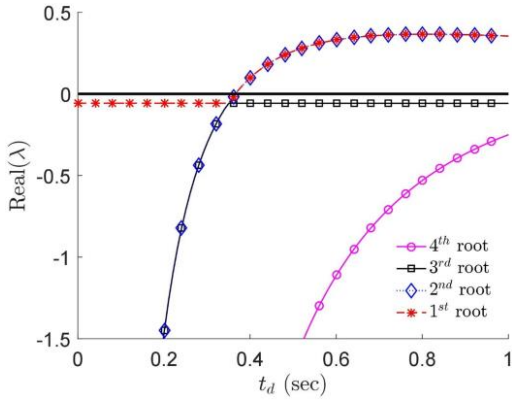

Fig. 9. Variation of first four characteristic roots of (16) with respect to $t_d$.

Smith predictor (ASP) with DFR series approximation of time-delay and PI controller. Uncertain delays are induced in the network via a code running in intermediate node. The main objective is to stabilize the closed system for delays greater than $t_{crit}$ by implementing the proposed adaptive Smith predictor developed in this work. In practice there is a certain operating bounds for current and voltage associated with actuators. Any voltage/current greater than its operating value might damage the actuator or lead to undesirable functioning. Therefore, it is a good practice to maintain the voltage/current supply to actuators less than its nameplate specifications. Therefore, it needs to be shown that with the proposed scheme of adaptive Smith predictor, the control input is maintained below the actuator operating voltage (5 V). A 100% duty ratio corresponds to 5 V, whereas a 0% duty ratio is to 0 V.

Fig. 10 shows experimental results for time-delay variations from 0.03 s to 0.130 s. Time-delays in Fig. 10(a) is less than $t_{crit}$ and the performance of CSP and ASP schemes are similar. It should be noted that the maximum value attained by the control input signal is only about 57% duty ratio in case of both CSP and ASP schemes. Thus, for the delays less than $t_{crit}$, the performance of the CSP and ASP schemes are satisfactory.

Fig. 11 shows the system response for classical Smith predictor with a fixed time-delay of 0.06s and PI controller. The delay varies from 0.293 s to 0.389 s which is slightly greater than $t_{crit}$. Fig. 11(b) shows the response of the classical Smith predictor and the control input signal. The system response exhibits oscillations and does not accurately follow the set-point. The control signal rises to a value close to maximum (98 % duty ratio) indicating actuator saturation, and no satisfactory performance is observed for a large mismatch in the nominal fixed delay and actual delay in the network.

Fig. 12(b) shows the response of the proposed adaptive Smith predictor developed with DFR approximation of time-delay. The variation of time-delay is in the range of 0.370 s to 0.636 s which is approximately 1.6 times greater than $t_{crit}$. It can be seen from Fig. 12(b) that the closed-loop system performance is satisfactory with no oscillations in the system response. Also, the control signal attains a maximum value of about 87% duty ratio and a minimum value of about 20% duty ratio, thereby avoiding actuator saturation. The proposed scheme effectively mitigates the effects of large and random time-delays ($> t_{crit}$) in the network and improves system stability.

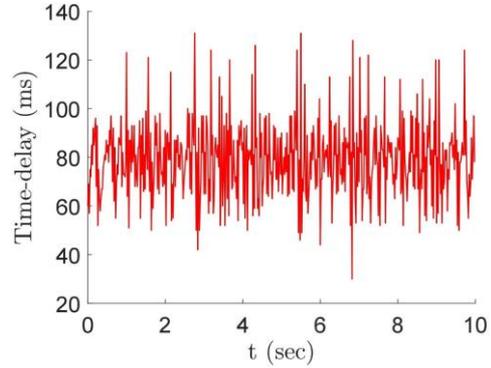

(a)

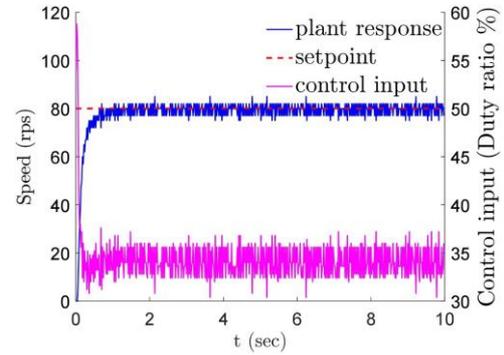

(b)

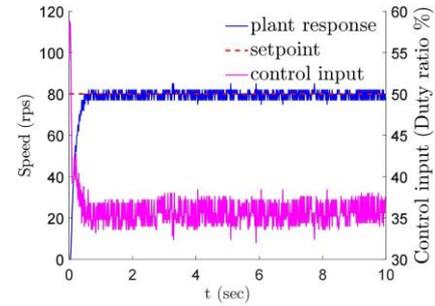

(c)

Fig. 10. (b) classical Smith predictor ($t_m = 0.06$s) with PI control; (c) adaptive Smith predictor with DFR series approximation.

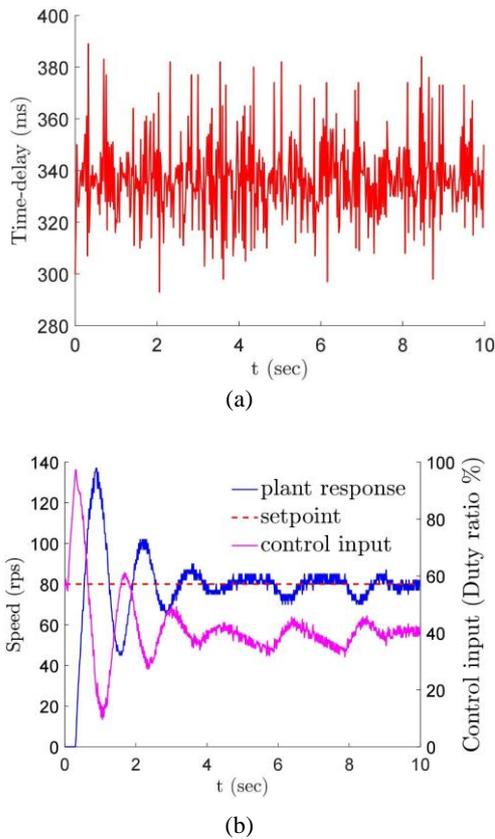

Fig. 11. With classical Smith predictor ($t_m = 0.06$ s) with PI controller.

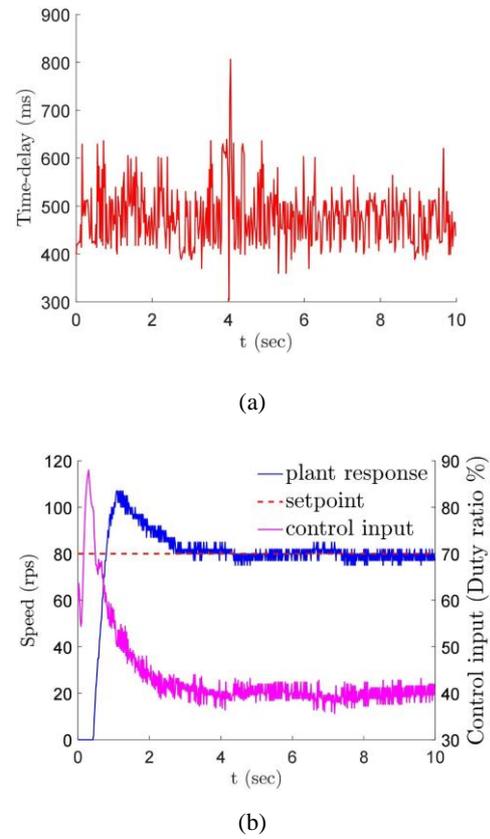

Fig. 12. With adaptive Smith predictor with DFR series approximation.

## 7. ALGORITHMIC STEPS

Next, the algorithm for delay estimation and delay compensation in conjunction with a discrete PI controller for DC servo motor speed regulation is presented.

**Remark 1:** The sampling time of 1 sec is considered in [10] whereas in the proposed work it is 0.02 sec due to the nature of application chosen. Maximum delay variation in [1] is found to be 600 msec for a separation of around 4 Km whereas in the proposed work we have attained a maximum delay of 800 msec for a distance of 60 meters. The nature of delay variation in [10] and [1] seems to be constant over some sampling time, whereas there are variations within a certain bound, at every sampling instant as seen from Figs. [10-12].

**Remark 2:** In [1, 10] the algorithms for control are implemented through PC with high-performance math processing software, whereas all the algorithms (delay estimation, delay compensation, and controller) are implemented in embedded platform with limited computing capabilities, and so [10] and [1] significantly differ in both theoretical and practical aspects from the present work.

**Remark 3:** The convergence of the proposed WNCS can be studied from $e_2 = r - \hat{y} - (y_d - \hat{y}_d)$ (13). It is required that $r - \hat{y} \to 0$ and $y_d - \hat{y}_d \to 0$, hence the error term $e_2 \to 0$. The convergence of the term $y_d - \hat{y}_d$ is discussed in Section 4.1.4. The error term $r - \hat{y} \to 0$ if the nominal feedback $\hat{y}$ is a good estimate of the non delayed feedback $y$. Hence, if $\hat{G}(z)$ and $G_{dm}(z)$ are proper, $e_2 \to 0$ in finite time. Figs. [10-12] describe that the plant response with the proposed adaptive Smith predictor closely follows the set-point. Therefore, it can be said that the error $e_2 \to 0$ in finite time for the developed WNCS. The convergence speed can be improved by starting the Smith predictor with an initial delay [1]. The detailed convergence analysis of the developed WNCS with proofs are left for another publication due to length restrictions here.

## 8. CONCLUSION AND FUTURE SCOPE

The present work deals with embedded control system development for DC servo motor over Bluetooth wireless network. The prototype is built with limited computing resource, thereby innovating the computational implementation in terms of delay estimation and adaptive Smith predictor, thereby preventing serious performance degradation due to uncertain time-delay. Stability analysis is provided to validate the effect of delay. Experimental results

**Algorithm 1**

1: Set PI parameters $I_{thres}$, $I_{sum} = 0$, $K_P$, $K_I$.
2: Set ASP parameters $x[n-1] = x[n-2] = x[n-3] = 0, y[n-1] = y[n-2] = y[n-3] = 0$.
3: **while** *Serial.available*() **do**
4:     Receive feedback signal ($y_d$).
5:     Register the receive time stamp ($t_{2pr}$).
6:     Estimate $t_m$ using rules in Section 3.1 and (4).
7:     Compute $e_1 = r - y_d$, $e_2 = e_1 - y$.
8:     **procedure** DRIVE = PI CONTROLLER(PI parameters)
9:        P = $e_2 \times K_P$,    I = $I_{sum} \times K_I$.
10:        DRIVE = P + I, $I_{sum} = I_{sum} + e_2$.
11:        **if** DRIVE > actuator saturation **then**
12:           set DRIVE = *constraint*(Drive, 0, 255).
13:        **if** $I_{sum} > I_{thres}$ **then**
14:           set $I_{sum} = I_{thres}$.
15:        **return** DRIVE
16:     Send DRIVE signal.
17:     Register sent time stamp ($t_{1pr}$).
18:     **procedure** $y$ = ASP(ASP parameters)
19:        Set $x[n]$ = DRIVE.
20:        Compute (12).
21:        **return** $y$

validate the proposed implementation scheme by showing that the proposed scheme can stabilize the closed-loop system for delays greater than $t_{crit}$ under highly uncertain time-delay variation situation.

Future scope of this research includes implementation of the said methodology in a loaded network, investigation of data loss issues, appropriate algorithms for inherent nonlinear attributes in the plant, and implementation for a real-time artificial pancreas system (APS).

## REFERENCES

[1] C. L. Lai, and P. L. Hsu, "Design the remote control system with the time-delay estimator and the adaptive Smith predictor," *IEEE Transactions on Industrial Informatics*, vol. 6, no. 1, pp. 73-80, 2010.

[2] doi J. Eker, A. Cervin, and A. Hörjel, "Distributed wireless control using Bluetooth," *Proceedings of the IFAC Conference on New Technologies for Computer Control*, 2001.10.1016/S1474-6670(17)32965-8

[3] N. J. Ploplys, P. A. Kawka, and A. G. Alleyne, "Closed-loop control over wireless networks," *IEEE Control Systems*, vol. 24, no. 3, pp. 58-71, 2004.

[4] C. H. Chen, C. L. Lin, and T. S. Hwang, "Stability of networked control systems with time-varying delays," *IEEE Communications Letters*, vol. 11, no. 3, 2007.

[5] J. D. Boeij, M. Haazen, P. Smulders, and E. Lomonova, "Low-latency wireless data transfer for motion control," *Journal of Control Science and Engineering*, vol. 2009, Article ID 591506, 2009.

[6] A. Hernandez, *Wireless Process Control Using IEEE 802.15. 4 Protocol*, Master Degree Thesis, Royal Institute of Technology, 2010.

[7] D. Yue, Q. L. Han, and J. Lam, "Robust H-inf control for uncertain networked control systems," 2004.

[8] G. Q. Qiu, D. M. Cao, S. Liu, and J. J. Bao, "GPC control strategy and simulation of a ZigBee based wireless networked control system," *Applied Mechanics and Materials*, vols. 687-691, pp. 641-644, 2014.

[9] Y. Wu, Y. Wu, and Y. Zhao, "Mode-dependent controller design for networked control system with average dwell time switching," *Transactions of the Institute of Measurement and Control*, vol. 39, no. 10, pp. 1577-1589, 2017.

[10] C. Suryendu, S. Ghosh, and B. Subudhi, "Variable gain output feedback control of a networked temperature control system based on online delay estimation," *Asian Journal of Control*, vol. 19, no. 3, pp. 1250-1254, 2017.

[11] C. P. Vyasarayani, S. Subhash, and T. Kalmár-Nagy, "Spectral approximations for characteristic roots of delay differential equations," *International Journal of Dynamics and Control* vol. 2, no. 2, pp. 126-132, 2014.

[12] R. Wang, G. P. Liu, W. Wang, D. Rees, and Y. B. Zhao, "Guaranteed cost control for networked control systems based on an improved predictive control method," *IEEE Transactions on Control Systems Technology*, vol. 18, no. 5, pp. 1226-1232, 2010.

[13] A. Onat, A. Teoman Naskali, and E. Parlakay, "Model based predictive networked control systems," *IFAC Proceedings*, vol. 41, no. 2, pp. 13000-13005, 2008.

[14] G. P. Liu, Y. Xia, D. Rees, and W. Hu, "Design and stability criteria of networked predictive control systems with random network delay in the feedback channel," *IEEE Transactions on Systems, Man, and Cybernetics, Part C*, vol. 37, no. 2, pp. 173-184, 2007.

[15] A. Ulusoy, G. Ozgur, and A. Onat, "Wireless model-based predictive networked control system over cooperative wireless network," *IEEE Transactions on Industrial Informatics*, vol. 7, no. 1, pp. 41-51, 2011.

[16] S. L. Du, X. M. Sun, and W. Wang, "Guaranteed cost control for uncertain networked control systems with predictive scheme," *IEEE Transactions on Automation Science and Engineering*, vol. 11, no. 3, pp. 740-748, 2014.

[17] L. Huang and Y. Fang, "Convergence analysis of wireless remote iterative learning control systems with dropout compensation," *Mathematical Problems in Engineering*, vol. 2013, Article ID 609284, 2013.

[18] R. Patael, D. Deb, R. Dey, and V. E. Balas, "Model reference adaptive control of microbial fuel cells," *Adaptive and Intelligent Control of Microbial Fuel Cells*, Springer, Cham, pp. 109-121, 2019.

[19] S. He, J. Song, and F. Liu, "Robust finite-time bounded controller design of time-delay conic nonlinear systems using sliding mode control strategy," *IEEE Transactions on Systems, Man, and Cybernetics: Systems*, vol. 49, no. 11, pp. 1863-1873, Nov. 2018.

[20] S. He, Q. Ai, C. Ren, J. Dong, and F. Liu, "Finite-time resilient controller design of a class of uncertain nonlinear systems with time-delays under asynchronous switching," *IEEE Transactions on Systems, Man, and Cybernetics: Systems*, vol. 49, no. 2, pp. 281-286, Feb. 2019.

[21] S. Shi, S. Xu, W. Liu, and B. Zhang, "Global fixed-time consensus tracking of nonlinear uncertain multiagent systems with high-order dynamics," *IEEE Transactions on Cybernetics*, 2018.

[22] H. Shen, F. Li, S. Xu, and V. Sreeram, "Slow state variables feedback stabilization for semi-Markov jump systems with singular perturbations," *IEEE Transactions on Automatic Control*, vol. 63, no. 8, pp. 2709-2714, 2018.

[23] B. Subudhi, S. Bonala, S. Ghosh, and R. Dey, "Robust analysis of networked control systems with time-varying delays," *IFAC Proceedings Volumes* vol. 45, no. 13, pp. 75-78, 2012.

[24] R. Datta, R. Dey, B. Bhattacharya, and A. Chakrabarti, "Delayed state feedback controller design for inverted pendulum using TS fuzzy modeling: an LMI approach," *Innovations in Infrastructure*, Springer, Singapore, pp. 67-79, 2018.

[25] M. Ghorbani and P. Bogdan, "A cyber-physical system approach to artificial pancreas design," *Proc. of International Conference on Hardware/Software Codesign and System Synthesis (CODES+ ISSS)*, IEEE, 2013.

[26] A. Nath, D. Deb, R. Dey, and S. Das, "Blood glucose regulation in type 1 diabetic patients: an adaptive parametric compensation control-based approach," *IET Systems Biology*, vol. 12, no. 5, p. 291, 2018.

[27] S. F. Ali, and R. Padhi, "Optimal blood glucose regulation of diabetic patients using single network adaptive critics," *Optimal Control Applications and Methods*, vol. 32, no. 2, pp. 196-214, 2011.

[28] M. Zhou, J. Ren, J. Qi, D. Niu, and G. Li, "Emerging technologies in knowledge discovery and data mining," *Lecture Notes in Computer Science*, pp.87-98, 2007.

[29] J. Lam, "Convergence of a class of Pade approximations for delay systems," *International Journal of Control*, vol. 52, no. 4, pp. 989-1008, 1990.

[30] C. P. Vyasarayani, "Galerkin approximations for higher order delay differential equations," *Journal of Computational and Nonlinear Dynamics*, vol. 7, no. 3, pp. 031004, 2012.

**Santosh Mohan Rajkumar** is currently working at Indian Oil Corporation Limited as an Operations Officer. He has obtained his Bachelor's degree in Electrical Engi- neering from National Institute of Tech- nology, Silchar, India in 2017. His re- search interest includes system identifica- tion, control of networked and distributed systems, robust control applications, embedded control applications, and time-delay sytems.

**Sayan Chakraborty** obtained his Bache- lor's degree in Electrical Engineering from National Institute of Technology, Silchar, India in 2017. He is currently pursuing a Master's degree in Systems and Control from the Dept. of Electrical Engineering at Indian Institute of Technology Hyderabad, India. His research interests include net- work control systems, embedded control, time-delay sytems, reduced order modelling, parameter identifi- cation, and artificial intelligence.

**Rajeeb Dey** is currently teaching in National Institute of Technology Silchar, As- sam, India as an Assistant Professor in the Department of Electrical Engineering. He has obtained his Masters degree from IIT Kharagpur and a Ph.D. from Jadavpur University, in 2007 and 2012, respectively both in Control System Engineering. He is a senior member of IEEE (CSS) since 2013 and a member of the Institution of Engineers (India). His research interest includes time-delay systems, control of biomed- ical systems, network control systems.

**Dipankar Deb** has completed his Ph.D. degree from University of Virginia, Char- lottesville with Prof. Gang Tao, IEEE Fel- low and Professor in the Department of ECE in 2007. He received his B.E. degree from NIT Karnataka, Surathkal (2000) and M.S. degree from the University of Florida, Gainesville (2004). In 2017, he was elected to be an IEEE Senior Member. He has served as a Lead Engineer at GE Global Research Ben- galuru (2012-15) and as an Assistant Professor in EE, IIT Guwa- hati 2010-12. Presently, he is a Professor in Electrical Engineer- ing at IITRAM Ahmedabad. He is also the Book Series Editor, Control System Series/CRC Press/Taylor and Francis Group. He is an Associate Editor of IEEE Access Journal. He has published a total of 17 SCI indexed Journal articles and holds 6 granted US patents.